\documentclass[pre,twocolumn,english,superscriptaddress,floatfix,longbibliography]{revtex4-2}

\usepackage{ulem}
\usepackage{graphicx}
\usepackage{dcolumn}
\usepackage{bm}
\usepackage{xcolor}
\usepackage{amsmath}
\usepackage{amssymb}
\usepackage{lineno}
\usepackage{hyperref}
\hypersetup{colorlinks=true, citecolor=blue, linkcolor=black, urlcolor=blue}

\newcommand{\ket}[1]{|#1\rangle}
\newcommand{\bra}[1]{\langle#1|}
\newcommand{\ccw}{\circlearrowleft}
\newcommand{\cw}{\circlearrowright}
\newcommand{\dbar}{\overline{D}}
\newcommand{\hcd}{\tilde{H}_\textsc{cd}}
\newcommand{\xf}{x_\mathrm{T}}
\newcommand{\ketbra}[2]{|#1\rangle\langle #2|}

\begin{document}

\title{Exploring the topology of a non-Hermitian superconducting qubit \\
using shortcuts to adiabaticity}

\author{Serra Erdamar}
\altaffiliation[These authors contributed equally to this work.]{}
\affiliation{Department of Electrical and Systems Engineering, Washington University, St. Louis, Missouri 63130, USA}
\affiliation{Department of Physics, Washington University, St. Louis, Missouri 63130, USA}
\author{Maryam Abbasi}
\altaffiliation[These authors contributed equally to this work.]{}
\author{Weijian Chen}
\affiliation{Department of Physics, Washington University, St. Louis, Missouri 63130, USA}
\author{Niklas H\"ornedal}
\author{Aur\'elia Chenu}
\affiliation{Department of Physics and Materials Science, University of Luxembourg, L-1511 Luxembourg, G. D. Luxembourg}
\author{Kater W. Murch}
\email{murch@physics.wustl.edu}
\affiliation{Department of Physics, Washington University, St. Louis, Missouri 63130, USA}

\date{\today}

             
\begin{abstract}
Open quantum systems described by a non-Hermitian Hamiltonian exhibit rich dynamics due to the topology of their complex energy spectrum. By encircling an exceptional point degeneracy, this topology allows for topological state transport, chiral geometric phases, and eigenvalue braiding. To access these topological features, it is desirable to drive the system adiabatically.  However, adiabatic transport in a system with complex spectrum is conventionally only possible for the eigenstate whose eigenenergy has the lowest loss.  Previous experiments have demonstrated such adiabatic evolution for the quantum state with relative gain, yet observed a breakdown in adiabaticity for quantum states with relative loss. In this work, we harness a shortcut to adiabaticity---counterdiabatic driving---to avoid the effects of loss while maintaining trajectories that follow the instantaneous eigenstates in significantly shorter timescales. We experimentally investigate the robustness of this control method using a superconducting transmon circuit with engineered dissipation. We observe that counterdiabatic driving stabilizes quasistatic transport and preserves the complex energy spectrum's topology.   
\end{abstract} 

\maketitle

\section{Introduction} 
The behavior of an open quantum system is typically modeled with a Liouvillian superoperator, which captures  the effects of a dissipative environment~\cite{Hatano2019,Minganti2019, Minganti2020}. Under some conditions, the dynamics can equivalently be cast in terms of a non-Hermitian Hamiltonian~\cite{Nagh19,chen21_jumps,Chen_22,Minganti2019,Kumar2021}. In either case, these generators of time translation often have complex spectra, with eigenenergy surfaces described by Riemann sheets. The complex energy surfaces therefore introduce the opportunity to control these systems based on the topology of their energy landscape~\cite{Xu2016, Holler2020,zhang2024,Guria2024,chavva25,Kumar2021,Pick2019,Milb15,Hass17,Berry11,Arkh24} using adiabatic tuning of the systems' Hamiltonian (or Liouvillian) parameters. Of particular interest are regions where the complex energy surfaces exhibit exceptional point degeneracies and branch cuts, leading to non-trivial state evolution.

\begin{figure}[]
    \includegraphics[width=1\linewidth]{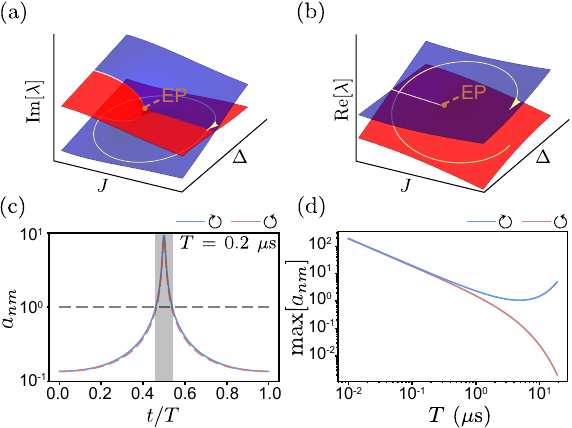}
    \caption{{\bf Complex energy spectra and adiabaticity of non-Hermitian systems near EP2.} (a, b) The imaginary and real parts of the PT-dimer eigenvalues correspond to Riemann surfaces. A path tuning the Hamiltonian parameters ($J$, $\Delta$) to encircle the EP in the clockwise direction is indicated in yellow. (c) Adiabaticity parameter [Eq.~(\ref{eq_adiabatic_cond})] for the parameter path with loop period $T=0.2\ \mu\mathrm{s}$. The shaded gray region highlights when $a_{nm}>1$. (d) $\mathrm{max}[a_{nm}]$ versus $T$ for counter-clockwise and clockwise encircling directions. The divergence of $\mathrm{max}[a_{nm}]$ at large $T$ is a consequence of gain/loss effects. }
    \label{fig1_schematic}
\end{figure}

This effect is illustrated in the paradigmatic example of the \textit{Parity-Time symmetry} (PT) dimer~\cite{Bender2005}. In this two-state system, one state experiences amplitude gain (at rate $\kappa$) while the other state features an equivalent level of loss. These two states, detuned in energy by $\Delta$, are coupled at a rate $J$. The non-Hermitian Hamiltonian of this model can be expressed as a $2\times 2$ matrix,
\begin{align}
H_\textsc{pt} = 
\begin{pmatrix}
\Delta-i \kappa&J\\
J&+i\kappa
\end{pmatrix}. \label{eq:ptdimer}
\end{align} 
$H_\textsc{pt}$ exhibits complex eigenvalues and a second-order exceptional point degeneracy (EP2) at $J=\kappa,\  \Delta=0$. Figure \ref{fig1_schematic}(a,b) displays the real and imaginary parts of the eigenvalues. The eigenvalues feature a rich topology that arises from the square-root dependence of the eigenvalues on the system parameters. We also illustrate in Fig.~\ref{fig1_schematic}(a,b) a potential parameter path (yellow curve) to be implemented. The diagram illustrates how, on such a path, the topology would cause one eigenstate to be adiabatically transported into the other by tuning the Hamiltonian parameters around the EP2. The consequences of such adiabatic evolution around an EP2 has been studied experimentally on a variety of platforms, including optomechanical systems~\cite{Xu2016,Guria2024}, optics~\cite{Dopp16,Li2020,Li2022}, NV centers in diamond~\cite{liu2021}, photonics~\cite{gao25}, nano-oscillators~\cite{Ho2024}, and superconducting qubits~\cite{Abb22}.

In these experiments, the topological effects exhibit a characteristic trajectory chirality when the system is transported around an exceptional point. The total gain or loss accumulated along a closed path depends on the specific trajectory taken in parameter space. Crucially, when the path encloses an EP2, reversing its direction leads to a reversal of the net gain or loss. This chirality directly affects the system’s dynamics.  When a state is transported around an EP2 following the eigenstate associated with gain, the evolution can remain adiabatic and follow the Riemann surface. In contrast, if the state follows the surface with loss, adiabaticity breaks down rapidly, as any small coupling to the gain state leads to exponential amplification~\cite{chen21_jumps,Abb22,Graefe2013}. These experiments usually require slow variation of system parameters. While not a significant issue in classical systems, in quantum systems, decoherence becomes significant for long evolution times.

In this work, we overcome these challenges by applying counterdiabatic driving to enhance the adiabatic response of a non-Hermitian system. Counterdiabatic driving~\cite{Berry2009,Demirplak2003,Chen2010,Bason2011,Jarzynski2013,delCampo2013,cep23,Cepaite2024,morawetz24,Deffner2014,An2016,Du2016,Zhang2013} is a control method that allows one to effectively achieve adiabatic response over shorter timescales~\cite{GuryOdelin2019,Torrontegui2013}. We experimentally implement this protocol on a non-Hermitian qubit formed from the energy levels of a dissipative superconducting circuit. The additional drives allow us to resolve dynamics consistent with the topology of the Riemann surfaces while mitigating non-adiabatic effects stemming from loss and faster-than-adiabatic parameter tuning. We characterize the efficacy of control in terms of how closely the resulting evolution follows the system's instantaneous eigenstates, and the preservation of the topology of the system's energy landscape.

The rest of this article is organized as follows: In Sec.~\ref{sec:adiabatic}, we review key features of adiabaticity in non-Hermitian systems. Section~\ref{sec:method} introduces the experimental platform and defines the Hamiltonian and its properties. Section~\ref{sec:encircle} reports experimental measurements of encircling EP2 and the resulting breakdown in adiabaticity. We then present the formalism for counterdiabatic driving in Sec.~\ref{sec:counterdiabatic} and demonstrate its efficacy. In Sec. \ref{sec:NH_counterdiabatic}, we show the conditions to preserve a Hermitian control and study the effect of anti-Hermitian contributions to the counterdiabatic Hamiltonian. In Sec.~\ref{sec:resolving}, we demonstrate that the counterdiabatic driving preserves the topology of the complex energy landscape. Section~\ref{sec:conclusion} concludes the manuscript and provides perspective for further work. 

\section{Adiabaticity in non-Hermitian dynamics} \label{sec:adiabatic}
We illustrate the challenges associated with adiabaticity in non-Hermitian systems with the PT-dimer Hamiltonian [Eq.~\eqref{eq:ptdimer}]. Because $H_\textsc{pt}$ is non-Hermitian, we have to take care in defining \textit{right} and \textit{left} eigenstates that form a biorthogonal basis. The right eigenstates are defined as $H_\textsc{pt}\ket{R_n}=\lambda_n\ket{R_n}$. Correspondingly, we denote the left eigenstates as $\bra{L_n}H_\textsc{pt}  =  \lambda_n \bra{L_n}$. The right and left eigenstates are biorthogonal partners, satisfying $\bra{L_n}R_m\rangle = \delta_{nm}$~\cite{Ibez2011, Ibez2014}.  When the parameters $J$ and $\Delta$ are tuned in time, such that they follow the path indicated in Fig.~\ref{fig1_schematic}(a,b), the associated eigenstates and eigenvalues  of $H_\textsc{pt}$ change correspondingly. The degree to which parameter changes are adiabatic can be quantified via the following condition for the adiabaticity parameter $a_{nm}$:
\begin{equation}
a_{nm} = \frac{|\langle L_n(t)|\partial_t R_m(t)\rangle|}{|\lambda_n(t) - \lambda_m(t)|}e^{-I_{nm}(t)} \ll 1,
\label{eq_adiabatic_cond}
\end{equation}
where $I_{nm}\equiv \mathrm{Im}\left[\int_{0}^{t} (\lambda_m(t')-\lambda_n(t'))dt'\right]$ and assuming no crossing of the imaginary parts~\cite{Ibez2014}. When $a_{nm} \ll 1$, transitions between the eigenstates $\ket{R_m}$ and $\ket{R_n}$ 
are suppressed. This guarantees that a system initialized in $\ket{R_m}$ will remain in that state, despite changes in the Hamiltonian. The two factors in Eq.~(\ref{eq_adiabatic_cond}) reveal two important aspects for adiabaticity: the first factor requires the evolution to be slow compared to the energy gap between the system's energy eigenstates, thus avoiding unwanted transitions. The second factor is unique to complex spectra and captures the effects of gain/loss that exponentially favor the state with relative gain.

Figure~\ref{fig1_schematic}(c) evaluates the adiabaticity condition (\ref{eq_adiabatic_cond}) in the PT-dimer as a function of time $t$, for a generic parameter path that encircles an EP2.  The loop starts at $J\gg \kappa$ and $\Delta=0$.  The control time is denoted by $T$. Near $t/T=0.5$, the energy gap is small compared to rate of the parameter change, leading to a breakdown in the adiabaticity condition. If the control time is increased, the effects of gain/loss start to become significant, as illustrated in Fig.~\ref{fig1_schematic}(d). There, we plot $\mathrm{max}[a_{nm}]$ for the same initial eigenstate, for parameter paths of different durations that encircle the EP in either direction. For small $T$, $\mathrm{max}[a_{nm}]$ is dominated by the first factor in Eq.~(\ref{eq_adiabatic_cond}), which decreases as $T^{-1}$. However, for large $T$, $\mathrm{max}[a_{nm}]$ differs for counter-clockwise or clockwise directions. Correspondingly, one path corresponds to overall relative ``gain'' (pink), and the other to relative ``loss'' (light blue). The loss path has a dramatic breakdown in adiabaticity as the gain/loss factor becomes significant~\cite{Gilary2013,Uzdin2011}. This breakdown is ultimately responsible for the trajectory chirality observed in prior demonstrations of encircling.

\begin{figure}[t]
    \includegraphics[width=1\linewidth]{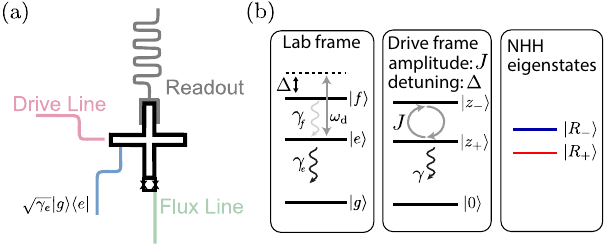}

    \caption{ {\bf Setup}.  (a) Sketch of the superconducting transmon circuit, drive line, dissipative channel, readout resonator, and fast flux line for frequency tuning.  (b) The lowest three energy levels of the transmon are labeled $\ket{g}$, $\ket{e}$, and $\ket{f}$. In the frame rotating with the drive of amplitude $J$ and detuning $\Delta$, the states are labeled  $\ket{0}$, $\ket{z_+}$, and $\ket{z_-}$. Dissipation promotes rapid decay of $\ket{z_+}$ to $\ket{0}$ and the dynamics are postselected on the $\{ \ket{z_+},\ket{z_-}\}$ submanifold. The eigenstates of the effective non-Hermitian Hamiltonian (NHH) system are specified by $\ket{R_\pm}$.}
    \label{fig_setup}
\end{figure}

\section{Setup}\label{sec:method}

As depicted in Fig.~\ref{fig_setup}(a), the system is based around a SQUID-tunable transmon circuit that is coupled to an off-chip dissipative channel used for frequency-selective dissipation. This dissipative channel is realized by a coaxial resonator with resonance frequency $\sim 4.25$~GHz. By applying a static flux bias to the SQUID, the lowest energy transition of the transmon can be tuned into resonance with the dissipative channel. The effect of this dissipative channel can be captured by a jump operator $\sqrt{\gamma_e}\ket{g}\bra{e}$. In contrast, the dissipation of the $\ket{f}$ state, given by $\sqrt{\gamma_f}\ket{e}\bra{f}$, with \(\gamma_f\) is comparatively small.  We define $4\kappa \equiv  \gamma = \gamma_e - \gamma_f = 1.16 \ \mu\mathrm{s}^{-1}$. The transmon is dispersively coupled to a microwave readout resonator ($f_\mathrm{ro} = 6.889$~GHz). The readout resonator enables multi-state, single-shot readout in the circuit's $\{\ket{g},\ket{e},\ket{f}\}$ basis (Fig.~\ref{fig_setup}b). We use this readout to perform postselection on the no-jump evolution that preserves the excited state manifold $\{\ket{e},\ket{f}\}$.  Finally, we drive the $\{\ket{e},\ket{f}\}$  transition with detuning $\Delta$ from its resonance at $f_\mathrm{q} = 4.095$~GHz.  The drive induces a coupling given by $J = J_x + i J_y = |J| e^{i \phi}$. In the frame rotating with this drive the new eigenbasis is denoted $\{\ket{z_+},\ket{z_-}\}$. 
The combination of dissipation, driving, and post-selection on the no-jump evolution leads to an effective non-Hermitian Hamiltonian, $H_\mathrm{eff} = E(\hat{I}+ \hat{\sigma}_z) + J_x \hat{\sigma}_x + J_y \hat{\sigma}_y$. This reads, in the  $\{\ket{z_+},\ket{z_-}\}$ basis, as 
\begin{equation} \label{eq_Heff}
H_\mathrm{eff}=\begin{pmatrix}
2E &J^*\\
J& 0
\end{pmatrix}
\end{equation}
with the complex energy $E = \Delta/2 - i \kappa$. 
$H_\mathrm{eff}$ is referred to as the passive PT-dimer Hamiltonian. It features an imbalance of loss between the two states, as opposed to the balanced gain and loss typically found in active PT-dimer systems, such as presented above. They are easily related through $H_\mathrm{eff} = H_\textsc{pt} -i \kappa \hat{I}$.

The eigenvalues of $H_\mathrm{eff}$ are $\lambda_{\pm}= E \pm \sqrt{|J|^2+E^2}$. 
When the parameters $J$ and $\Delta$ are varied, they allow exploring the topology of Riemann surfaces 
\footnote{Strickly speaking, having a Riemann surface would require the eigenvalues to be holomorphic, which is not the case here. We thus rather deal with a generalization of a Riemann surface, known as a `branched covering'. This does not impact the topological features.}
depicted in Fig.~\ref{fig1_schematic}(a,b). The EP2s occur at $J=\pm \kappa$ and zero detuning.

The underlying chiral symmetry of $H_\mathrm{eff}$ allows us to parameterize the eigenvectors in terms of a complex mixing angle $\alpha = \alpha_\textsc{r} + i \alpha_\textsc{i}$ defined by $\tan(\alpha) = \frac{|J|}{E}$ (See App.~\ref{appA}). The limit of a real mixing angle ($\alpha_\textsc{i}=0$) corresponds to orthogonal eigenvectors while the \textit{hyperbolic angle} $\alpha_\textsc{i}$  quantifies the deviation from orthogonality. The right eigenvectors, defined from $(H_\mathrm{eff} - \lambda_{\pm})\ket{R_\pm}= 0$, can be expressed in terms of $\alpha$ and $\phi$ as $\ket{R_\pm} = e^{\mp i \frac{\phi}{2}}\cos(\alpha/2)\ket{z_\pm} \pm e^{\pm i \frac{\phi}{2}}\sin(\alpha/2)\ket{z_\mp}$.
In turn, the left eigenvectors,  that satisfy $\bra{L_\pm}(H_\mathrm{eff} - \lambda_{\pm})= 0$, are given by $\bra{L_\pm} = e^{\pm i \frac{\phi}{2}}\cos(\alpha/2)\bra{z_\pm} \pm e^{\mp i \frac{\phi}{2}}\sin(\alpha/2)\bra{z_\mp}$. The right and left eigenvectors form a bi-orthogonal basis, since $\langle{L_n}|{R_m} \rangle = \delta_{nm}$. 
The degree of non-orthogonality is reflected in the norms $\langle R_{\pm}|   R_\pm \rangle =   \langle L_{\pm}|   L_\pm \rangle=\cosh\alpha_\textsc{i}$, and in the overlap $\langle R_{-}|   R_+ \rangle = \langle L_{+}|   L_- \rangle = i\sinh\alpha_\textsc{i}$, both of which approach their orthogonal values as $\alpha_\textsc{i}\rightarrow 0$.

Since the biorthonormal eigenbasis can be expressed in terms of $\alpha$, the states can be obtained as the result of a complex rotation $\hat{C}_y(\alpha)$ about the $y$-axis of the Bloch sphere, applied to the $\{|z_+\rangle, |z_-\rangle\}$--basis, followed by a real rotation $\hat{R}_z(\phi) = e^{-i \frac{\phi}{2} \hat{\sigma}_z}$ that accounts for complex couplings:
\(\ket{R_\pm} = \hat{R}_z(\phi)\hat{C}_y(\alpha) \ket{z_\pm}\).
The complex rotation can itself be decomposed into a rotation $\hat{R}_y$ of angle $\alpha_\textsc{r}$ and boost $\hat{B}_y$ of hyberbolic angle $\alpha_\textsc{i}$ around the $y$ axis:
\begin{equation}
\hat{C}_y (\alpha) \equiv e^{ -i \frac{\alpha}{2} \hat{\sigma}_y} = e^{-i \frac{\alpha_\textsc{r}}{2} \hat{\sigma}_y }  e^{\frac{\alpha_\textsc{i}}{2}  \hat{\sigma}_y } \equiv \hat{R}_y(\alpha_\textsc{r}) \hat{B}_y(\alpha_\textsc{i}).
\end{equation}
The boost gives a correction to the eigenstates in terms of hyberbolic functions that depend only on the  imaginary contribution of the complex angle. 
The left eigenvalues are given analogously, $\bra{L_\pm} = \bra{z_\pm} \hat{C}_y(-\alpha)\hat{R}_z(-\phi)$. 
\section{Encircling the EP}\label{sec:encircle}

\begin{figure}[t]
    \includegraphics[width=1\linewidth]{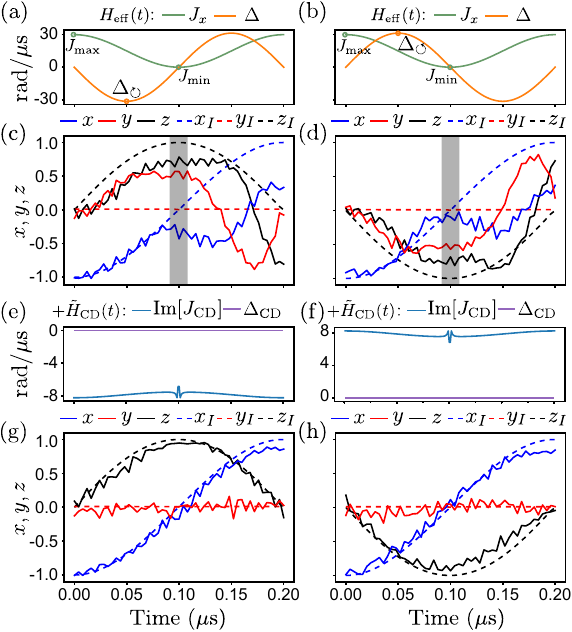}
    
    \caption{ {\bf Quantum state tomography of EP2 encircling  with and without counterdiabatic driving.} (a,b) The time dependence of the parameters in $H_\mathrm{eff}$ for $\Delta_\cw$ and $\Delta_\ccw$.  (c,d) Experimentally measured Pauli expectation values, $\{x, y, z\}$, plotted versus $t$ (solid lines) for $T=0.2\ \mu$s evolved with $H_\mathrm{eff}$. The dashed lines show the calculated Pauli expectation values for the instantaneous eigenstates $\{x_\mathrm{I}, y_\mathrm{I}, z_\mathrm{I}\}$. After $t=0.1\ \mu$s, the measured state deviates significantly from the instantaneous eigenstates, indicating a breakdown in adiabaticity. This corresponds to the region when $a_{nm}>1$ (gray box). (e,f) The time dependence of the parameters in $\hcd$ for $\Delta_\cw$ and $\Delta_\ccw$. (g,h) With the addition of counterdiabatic driving, the tomography reveals trajectories that closely follow the instantaneous eigenstates of $H_\mathrm{eff}$. The decay values for this experiment are $\gamma_e = 1.37~\mu\mathrm{s}^{-1}$ and $\gamma_f = 0.21~\mu\mathrm{s}^{-1}$, with $\kappa = 0.29~\mu\mathrm{s}^{-1}$.}
    \label{fig_time}
\end{figure}
We now study the evolution of the system as the parameters $J$ and $\Delta$ are tuned in time to encircle the EP2 at $J = \kappa$ and $\Delta=0$. The parameter paths, depicted in Fig.~\ref{fig_time}(a,b), are defined by the parameters $J_\mathrm{min}$, $J_\mathrm{max}$, and $\Delta_{\cw,\ccw}$ as
\begin{equation}
\begin{aligned}
    J(t) &= \frac{J_{\max}-J_{\min}}{2}\cos\left(\frac{2\pi t}{T}\right)+\frac{J_{\max}+J_{\min}}{2},
    \\
    \Delta(t) &= \Delta_{\cw,\ccw}\sin\left(\frac{2\pi t}{T}\right),
\end{aligned}
\label{eq:controls}
\end{equation}
where $\Delta_\cw <0$ defines a clockwise encircling direction and $\Delta_\ccw >0$ a counter-clockwise direction. We initialize the system in the eigenstate $\ket{R_-}$  of $H_\mathrm{eff}$, with $J=30\ \mathrm{rad}/ \mu\mathrm{s}$ and $\Delta=0$. We then tune $J$ and $\Delta$ as in Eqs.~(\ref{eq:controls}) with $J_\mathrm{max} = 30\ \mathrm{rad}/ \mu\mathrm{s}$, $J_\mathrm{min} = 0\ \mathrm{rad}/ \mu\mathrm{s}$ and $\Delta_{\cw} =  - 10 \pi\ \mathrm{rad}/ \mu\mathrm{s}$. The total evolution time $T$ is divided into $N=51$ time steps of size $\delta t = T/(N-1)$. We perform a set of successively longer duration experiments with  evolution times $t_{n+1} = t_n + \delta t$. For each evolution time, we perform quantum state tomography. We set $J=0$ and perform rotations in the $\{\ket{z_+},\ket{z_-}\}$ basis, followed by state readout. 
The Pauli expectation values are denoted $\{ x\equiv \langle{\hat{\sigma}_x}\rangle, y\equiv \langle{\hat{\sigma}_y}\rangle, z\equiv \langle{\hat{\sigma}_z}\rangle \}$. For a loop period $T = 0.2\ \mu \mathrm{s}$, the quantum state tomography is shown in Fig.~\ref{fig_time}(c). The solid lines represent the tomography results and the dashed lines are the Pauli components of the instantaneous eigenstates of $H_\mathrm{eff}$: 
$x_\mathrm{I}\equiv\bra{R_n}\hat{\sigma}_x\ket{R_n}$ 
with $n\to -$ for $t<T/2$ and $n\to +$ for $t>T/2$ (and similarly for $y_\mathrm{I}$ and $z_\mathrm{I}$). We observe that for $t<0.1 \ \mu\mathrm{s}$ the tomography results are in reasonable agreement with the instantaneous eigenstates, but for $t>0.1\ \mu\mathrm{s}$ the two differ significantly.   The divergence from the instantaneous eigenstates is most significant near the midpoint of the loop where the adiabaticity parameter $a_{nm}$ becomes comparable to unity, violating Eq.~(\ref{eq_adiabatic_cond}). This region is highlighted in the gray-shaded area, also see Fig.~\ref{fig1_schematic}. Figure~\ref{fig_time}(d) displays similar measurements for $\Delta_\ccw = + 10 \pi\ \mathrm{rad}/ \mu\mathrm{s}$. For both control loops, the observed breakdown of adiabaticity stems from the rapid parameter variation i.e.~from the first factor in Eq.~(\ref{eq_adiabatic_cond}).

\section{Counterdiabatic driving}\label{sec:counterdiabatic}

To avoid the non-adiabatic response observed in Fig.~\ref{fig_time}(c,d), we can employ the strategy of shortcuts to adiabaticity (STA). 
The goal of STA is to track the eigenstates of a system whose parameters change in time, and to do so in a finite time, without being limited by the adiabatic dynamics. 

Let us first consider the case of real coupling ($J = J_x$) for simplicity. The general case of complex coupling ($\phi\neq 0$) is given in App.~\ref{Counterdiabatic_Hamiltonian_phi}. Because the right eigenstates are obtained from the complex rotation $\ket{R_\pm} = \hat{C}_y(\alpha)\ket{z_\pm}$, it is straightforward to write dynamics giving such eigenstates at time $t$ starting from an eigenstate at $t_0$ in terms of a transport operator:  $\hat{T}(t- t_0) = \hat{C}_y(\alpha_t)\hat{C}_y(-\alpha_{t_0})=e^{-\frac{i}{2}(\alpha_t - \alpha_{t_0})\hat{\sigma}_y}$. This transport operator is also defined from the counterdiabatic Hamiltonian as $\hat{T}(t,t_0) \equiv e^{-i\int_{t_0}^t H_\textsc{cd}(s)ds}$. This leads to a simple expression of the counterdiabatic Hamiltonian 
\begin{equation} \label{eq:hcd}
   H_\textsc{cd}(t) = \frac{\dot{\alpha}_t}{2}\hat{\sigma}_y.
\end{equation}
Since $\alpha$ is complex, this control is generally non-Hermitian. 

Let us relate this simple derivation to the general form of the counterdiabatic driving Hamiltonian, derived in ~\cite{Ibez2011} as 
\begin{multline}
    H_\textsc{cd}(t) = i\sum_{n\in\pm} \ketbra{\partial_t R_n(t)}{L_n(t)}\\
    -\langle L_n(t)|\partial_t R_n(t)\rangle\ketbra{R_n(t}{L_n(t)}.
    \label{eq_HCD}
\end{multline}
$H_\textsc{cd}(t)$ comprises additional driving that is applied to the qubit as the parameters in $H_\mathrm{eff}$ are tuned.  The first term in Eq.~(\ref{eq_HCD}) negates the non-adiabatic transitions experienced by the states when the Hamiltonian parameters are tuned in time.  The second term is related to the geometry of the parameter space of adiabatic dynamics~\cite{Berry1984}, and is the integrand of the Berry phase. Therefore, adding $H_\textsc{cd}$ to the time-dependent Hamiltonian cancels out the non-adiabatic transitions~\cite{Cepaite2024} and preserves the system's geometric properties.  Here, the eigenvectors evolve (for $\phi=0$) as $\ket{\partial_t R_\pm} =  -i\frac{\dot{\alpha}_t}{2}\hat{\sigma}_y \ket{R_\pm}$, which is also equal to $\frac{\dot{\alpha}_t}{2}\ket{R_\mp}$. This second expression shows parallel transport, i.e. $\langle{L_\pm} | \partial_t R_\pm \rangle =0$, which removes the second line in Eq.~\eqref{eq_HCD}, while the first expression shows that $H_\textsc{cd}$ simplifies to Eq.~\eqref{eq:hcd}.

We decompose $H_\textsc{cd}$ into Hermitian and anti-Hermitian components $H_\textsc{cd}(t) = H_\textsc{cd}^{(\textsc{h})} + H_\textsc{cd}^{(\textsc{ah})} = \begin{pmatrix}
    0&J_\textsc{cd}\\
    -J_\textsc{cd}& 0
\end{pmatrix}$, with $J_\textsc{cd} = \frac{\dot{\alpha}_\textsc{i}}{2} - i\frac{\dot{\alpha}_\textsc{r}}{2}$. 
The Hermitian component $H_\textsc{cd}^{(\textsc{h})}  = \frac{1}{2}(H_\textsc{cd}+H_\textsc{cd}^\dagger)$ corresponds to additional drives on the qubit, which can be easy to implement experimentally. 
In contrast, the anti-Hermitian component $H_\textsc{cd}^{(\textsc{ah})} = \frac{1}{2}(H_\textsc{cd}-H_\textsc{cd}^\dagger)$,  arising from changes in the hyperbolic angle, can be difficult to implement in practice. We show in Sec.~\ref{sec:NH_counterdiabatic} that this part can only be eliminated on carefully chosen parameter paths that follow particular \textit{Apollonius circles}.

The parameter paths employed in Fig.~\ref{fig_time}(a,b) are in fact already quite close go Apollonius circles, with negligible anti-Hermitian component. In general one can implement an \textit{approximate} $\hcd = H_\textsc{cd}^{(\textsc{h})}$. 
We write $\hcd =\Delta_\textsc{cd}(t)|z_+\rangle\langle z_+| + \mathrm{Im}[J_\textsc{cd}(t)]\hat{\sigma}_y$ and display the parameters $\Delta_\textsc{cd}(t)$ and $\mathrm{Im}[J_\textsc{cd}(t)]$ in Fig.~\ref{fig_time}(e,f).
We now repeat the experiment with additional drives constituting $\hcd$ and use quantum state tomography to evaluate the effect of counterdiabatic driving. Figure~\ref{fig_time}(g,h) displays the results. We observe that the state closely follows the instantaneous eigenstates for both $\Delta_\cw$ and $\Delta_\ccw$. In particular, both directions show the expected quasistatic evolution~\cite{zhang2024}, i.e. $\ket{R_-}\to\ket{R_+}$. 

To quantify the efficacy of the counterdiabatic driving, we use the average trace distance $\dbar$ to measure how closely the system follows the instantaneous eigenstates. We define $\rho_\mathrm{I} = \frac{1}{2}(\hat{I} + x_\mathrm{I} \hat{\sigma}_x + y_\mathrm{I} \hat{\sigma}_y + z_\mathrm{I} \hat{\sigma}_z)$ as the density matrix of the instantaneous eigenstates of $H_\mathrm{eff}$. $\rho_\mathrm{q}$ is the experimentally reconstructed density matrix, $\rho_\mathrm{q} = \frac{1}{2}(\hat{I} + x \hat{\sigma}_x + y \hat{\sigma}_y + z \hat{\sigma}_z)$. The trace distance is given as
\begin{equation}
\label{eq_trdist}
    D(\rho_\mathrm{I},\rho_\mathrm{q}) = \frac{1}{2}\mathrm{Tr}[\sqrt{(\rho_\mathrm{I}-\rho_\mathrm{q})^\dagger (\rho_\mathrm{I}-\rho_\mathrm{q})}].
\end{equation}
The average trace distance, $\dbar$ is given by time-averaging $D(\rho_\mathrm{I},\rho_\mathrm{q})$ over the entire trajectory. Applying this measure to the results in Fig.~\ref{fig_time}(c,d), we obtain $\dbar = 0.411$ and $0.378$, for the $\Delta_\cw$ and $\Delta_\ccw$ parameter paths respectively. 
When the counterdiabatic driving is included [Fig.~\ref{fig_time}(g,h)], $\dbar$ is reduced to $0.086$ and $0.067$, respectively.

\begin{figure}[t]
    \includegraphics[width=1\linewidth]{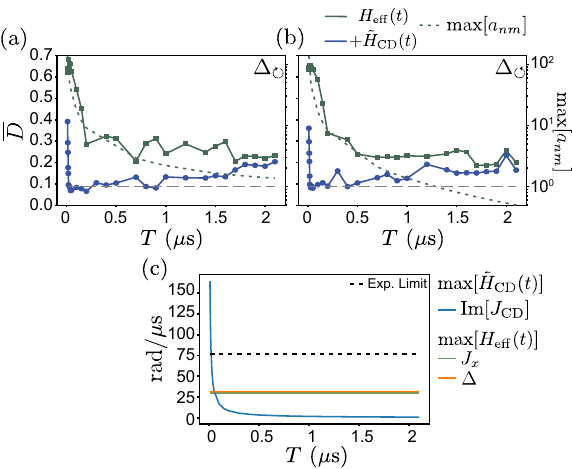}
    
    \caption{{\bf Evaluating the efficacy of counterdiabatic driving.} The Hamiltonian variation parameters are kept constant with $J_\mathrm{max} = 30\ \mathrm{rad}/ \mu\mathrm{s}$ and $J_\mathrm{min} = 0\ \mathrm{rad}/ \mu\mathrm{s}$.  $\dbar$ is evaluated for loops of different $T$ for each encircling direction (a) $\Delta_\cw$ and (b) $\Delta_\ccw$. Green squares correspond to $\dbar$ for evolution under $H_\mathrm{eff}$; blue circles correspond to evolution under $H_\mathrm{eff}+ \hcd$. $\mathrm{max}[a_{nm}]$ is plotted (right axis) as a function of $T$ (dashed line). The gray dashed line is when $\mathrm{max}[a_{nm}]=1$. (c) The maximum of each Hamiltonian drive parameter from $H_\mathrm{eff}(t)$ and $\hcd(t)$ plotted versus time $T$. The dashed black line is the experimental limitation on drive amplitude. The decay values for this experiment are $\gamma_e = 1.37~\mu\mathrm{s}^{-1}$ and $\gamma_f = 0.21~\mu\mathrm{s}^{-1}$, with $\kappa = 0.29~\mu\mathrm{s}^{-1}$.} 
    \label{fig_robust_param}
\end{figure}

We now turn to studying the effect of counterdiabatic driving versus the control time $T$: $\dbar$, for both $\Delta_\cw$ and $\Delta_\ccw$ is shown in Fig.~\ref{fig_robust_param}(a,b). The connected data points give $\dbar$ with  and without the control $\hcd$ (blue circles and green squares respectively). The dashed line (right axis) indicates $\mathrm{max}[a_{nm}]$.  For long periods ($T>1.5\ \mu$s) in \(\Delta_\ccw\), and $a_{nm}<1$ the evolution should approach adiabatic. Here, there is little difference observed in $\dbar\approx 0.2$ for evolution under $H_\mathrm{eff}+\hcd$ versus $H_\mathrm{eff}$ alone, because of quantum jumps and decoherence~\cite{chen21_jumps} taking place during the control protocol. For smaller $T$, we observe that the average trace distance for evolution under $H_\mathrm{eff}$ grows significantly, while  $\dbar$ for $H_\mathrm{eff}+\hcd$ decreases. Here, the addition of $\hcd$ allows the state to follow the instantaneous eigenstates. As $T\to 0$, we observe a sharp increase in $\dbar$, which can be explained by instrumentation limits: Fig.~\ref{fig_robust_param}(c) displays the maximum values of the parameters of $H_\mathrm{eff}$ and $\hcd$ versus $T$. For $T<0.02 \ \mu$s, the parameters in $\hcd$ become increasingly significant, ultimately reaching the limitations of the experimental apparatus as shown by the black dashed line in Fig.~\ref{fig_robust_param}(c). At this point, $\hcd$ cannot be experimentally implemented and we observe a  rapid increase in $\dbar$.

\section{Anti-Hermitian contribution to the counterdiabatic drive}\label{sec:NH_counterdiabatic}
\begin{figure}
    \centering
    \includegraphics[width=0.5\textwidth]{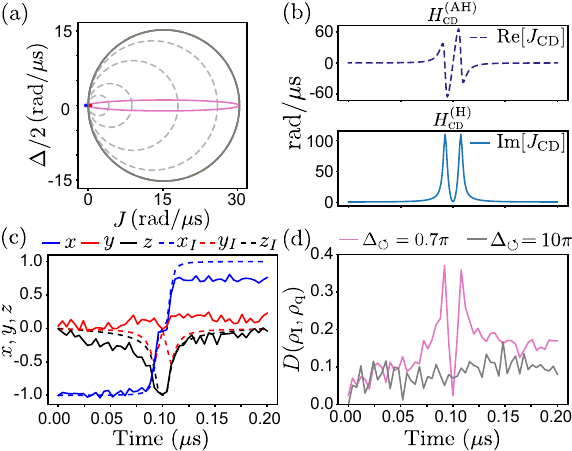}
    \caption{{\bf Effect of the anti-Hermitian component of the CD drive on the evolution.} (a) The gray solid line shows the Apollonius circle with $r = 0.9733$ and the inner dashed lines correspond to Apollonius circles with smaller radii. The pink trace crosses many circles, with $\Delta_\ccw=0.7\pi$, while $J_\mathrm{max}$ and $J_\mathrm{min}$ are kept constant. (b) The upper panel shows the required anti-Hermitian control components, $H_{\textsc{cd}}^{(\textsc{ah})}$ expressed in terms of $\textrm{Re}[J_\textsc{cd}]$ (purple trace). In the lower panel we plot the Hermitian control component, $H_{\textsc{cd}}^{(\textsc{h})}$, which is the implemented approximate counterdiabatic drive, $\hcd$, expressed as $\mathrm{Im}[J_\textsc{cd}]$ (blue trace). (c) The tomography result of the evolution corresponding to the pink ellipse in (a). (d) The trace distance for parameter paths that cross (pink) or follow (gray) Apollonius circles. The decay values for this experiment are $\gamma_e = 1.85~\mu\mathrm{s}^{-1}$ and $\gamma_f = 0.21~\mu\mathrm{s}^{-1}$, with $\kappa = 0.413~\mu\mathrm{s}^{-1}$.} 
    \label{fig_NH_contribution}
\end{figure}

We now look for the conditions for which $H_\textsc{cd}^{(\textsc{ah})}$ plays an important role in the counterdiabatic drive. We start by introducing the complex variable $\varepsilon \equiv \Delta/2 + iJ$ allowing us to view the parameter space as a complex plane. We can then rewrite the complex angle $\alpha$ as 
\begin{equation}\label{rotation angle}
\alpha(\varepsilon) = \frac{1}{2i}\log\left(\frac{\varepsilon-i \kappa}{\varepsilon^*-i \kappa}\right).
\end{equation}
This  gives the real and imaginary part of the angle as 
$\alpha_\textsc{r}(\varepsilon) = \frac{1}{2}\big(\arg(\varepsilon-i\kappa) + \arg(\varepsilon+i\kappa)\big)$ and 
$\alpha_\textsc{i}(\varepsilon) =  \frac{1}{2}\ln\left|\frac{\varepsilon+i\kappa}{\varepsilon-i\kappa}\right|$. 
 Importantly, the relevant parameters for the counterdiabatic drive are the derivatives of these angles. They vanish  on the set of points $\varepsilon$ satisfying 
\begin{subequations}
\begin{align}
\arg(\varepsilon - i\kappa) + \arg(\varepsilon + i\kappa) &= \text{constant}, \label{eq:cassini}\\ 
r\equiv \left| \frac{\varepsilon + i\kappa}{\varepsilon - i\kappa} \right|&= \text{constant}. \label{eq:apollonius}
\end{align}
\end{subequations}
This last equation \eqref{eq:apollonius} defines contours along which the ratio of distances from \(\varepsilon\) to the exceptional points \(\pm i\kappa\) remain constant. These contours, illustrated in Fig.~\ref{fig_NH_contribution}(a), are known as \textit{Apollonius circles}. Evolving along such a circle keeps \(\alpha_\textsc{i}(\varepsilon)\) constant, thus eliminating the anti-Hermitian component of the counterdiabatic drive.

We show in App.~\ref{appA} that  the angle between the two eigenstates $\ket{R_{\pm}}$ and their angle with the $y$-axis are equal and is given by $\theta\equiv \arccos|\tanh \alpha_\textsc{i}|$. This overlap angle depends solely on $\alpha_\textsc{i}$, so the contour lines of $\theta$ coincide with the Apollonius circles defined by the two EPs. Intuitively this is expected as only the hyperbolic part of the rotation can alter the overlap between eigenstates---the rotational part preserves it. This also explains why the counterdiabatic Hamiltonian can remain Hermitian when $\alpha_\textsc{i}$ is constant, since this is precisely when the overlap angle is constant.

We now illustrate how deviations from these special parameter paths lead to a breakdown in the effectiveness of the counterdiabatic drive when the anti-Hermitian component is neglected. 
We note that the parameter path eliminating the anti-Hermitian part of the counterdiabatic drive can be implemented by constructing a parametrized circle in the  ($J$, $\Delta/2$)-plane.  The Apollonius circle is centered at
\(c=( \kappa \frac{1 + r^2}{1 - r^2},0)\) with radius \( R=\frac{2 \kappa r}{|1 - r^2|}\),
where \( r = \left| \frac{\varepsilon + i\kappa}{\varepsilon - i\kappa} \right| \) determines the specific Apollonius circle. This results in \(J_{\max} = c+(R,0)\), \(J_{\min} = c-(R,0)\) and \(\Delta_\ccw = 2 R\).

Parameter paths that deviate from the Apollonius circle will in contrast require significant anti-Hermitian control components.  To study the effects of such deviation, we analyze a loop with the 
$J_{\max} = 30.3~\mathrm{rad}/\mu\mathrm{s}$, $J_\mathrm{min}=0.007~\mathrm{rad}/\mu\mathrm{s}$ and $\Delta_{\ccw} = 0.7 \pi ~\mathrm{rad}/\mu\mathrm{s}$, deviating from the Apollonius circle. As shown in Fig.~\ref{fig_NH_contribution}(a), the \(\alpha_\textsc{i}\) for this loop varies during the evolution as it crosses smaller Apollonius circles with different \(\alpha_\textsc{i}\). Consequently, the derivative of the anti-Hermitian part of \(\alpha\) is not constant and $H_\textsc{cd}$ acquires a significant anti-Hermitian component, especially near the center of the evolution path. The anti-Hermitian component of $H_\textsc{cd}$ is plotted in Fig.~\ref{fig_NH_contribution}(b) in dark purple in the upper panel. The imaginary part is plotted in blue in the lower panel and corresponds to a Hermitian drive. In Figure \ref{fig_NH_contribution}(c), we see the result of driving with approximate $\hcd$ on the pink path, which implements only the Hermitian drive, and the deviation of the qubit state from the instantaneous eigenstate. In particular $y$ deviates significantly from $y_\mathrm{I}$ as the loop approaches the EP. Figure~\ref{fig_NH_contribution}(d) displays $D$ versus time the trajectory. The pink trace pertains to a path that crosses many Apollonius circles and would require significant $H_\textsc{cd}^\textsc{(ah)}$. Since this control is not implemented, there is an increase in $D$ near the center of the loop. In comparison, the gray trace, corresponding to the near-Apollonius parameter path used in Fig.~\ref{fig_time}(c)(with $r = 0.9733$), maintains small values of $D$ along the whole trajectory. As we show in App.~\ref{appC}, the existence of such Hermitian-only counterdiabatic controls is related to underlying chiral symmetries of the Hamiltonian, and can be expected to occur in many other non-Hermitian systems.

\section{Resolving the Riemann topology with counterdiabatic driving}\label{sec:resolving}
\begin{figure}[t]
    \includegraphics[width=1\linewidth]{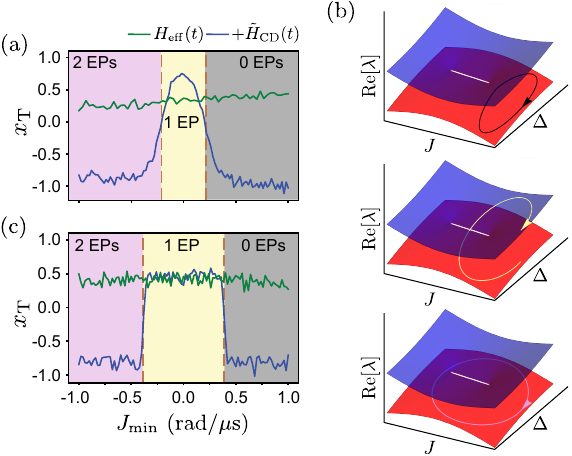}
    \caption{{\bf Resolving the Riemann topology.} (a) The $x$-component of the tomographically reconstructed state at the end of the parameter loop with period $T=0.2\ \mu \mathrm{s}$. The green curve is the result of driving with $H_\mathrm{eff}(t)$ and the blue curve is with the addition of $\hcd(t)$.  The loop parameters are $J_\mathrm{max} = 30\ \mathrm{rad}/ \mu\mathrm{s}$ and $\Delta_\mathrm{\cw} = -10\pi\  \mathrm{rad/}\mu\mathrm{s}$. $J_\mathrm{min}$ is swept from $-1$ to $1\ \mathrm{rad}/ \mu\mathrm{s}$. The vertical dashed lines represent the exceptional points at $J=\kappa$ with $\kappa = 0.21\ \mu\mathrm{s}^{-1}$.  (b) Sketches of the corresponding parameter paths for the three regions corresponding to encircling zero, one, or two EP2s.   (c) The $x$-component of the tomographically reconstructed state at the end of the parameter loop for $H_\mathrm{eff}(t)$ (green) and $H_\mathrm{eff}(t)+\hcd(t)$ (blue) with period $T=1\ \mu \mathrm{s}$ and $\kappa = 0.39\ \mu\mathrm{s}^{-1}$.}

    \label{fig_topo}
\end{figure}

We have so far demonstrated counterdiabatic driving for control loops that encircle a single EP2. To fully explore the Riemann sheet topology, we now change $J_\mathrm{min}$ so that the control can encircle zero, one, or two EP2s. We measure the tomography $x$-component evaluated at the end of the parameter path $\xf=x(t=T)$ with initial state $\ket{x_-}$. 
Figure \ref{fig_topo}(a) displays $\xf$ for different $J_\mathrm{min}$ with $\Delta_\cw$ and $T=0.2\ \mu$s. We first focus on the case where we employ counterdiabatic driving (blue curve). When the loop encircles zero EP2s (gray region), $x(t=T)\approx-1 = \langle x_- |\hat{\sigma}_x |x_-\rangle$; since no EP2 was encircled, there is no expected change in the state after one loop (as in Fig.~\ref{fig_topo}(b), top panel). When two EP2s are encircled (purple region), we observe similar behavior since the encircling does not pass through the branch cut and the state should remain on the same Riemann sheet (Fig.~\ref{fig_topo}(b), bottom panel). The yellow region corresponds to encircling one EP2, where the control parameters pass through a branch cut leading to a state change $\xf\approx  +1 = \langle x_+ |\hat{\sigma}_x |x_+\rangle$, as is shown in Fig.~\ref{fig_time}(g,h). We observe a sharp transition in the final state corresponding to where $J_\mathrm{min} = \pm \kappa$. In contrast, in the absence of counterdiabatic driving (green curve), $\xf$ does not reflect the features predicted by the Riemann topology. The starkly different values of $\xf$ for the three regions verify that the counterdiabatic driving resolves the fine features of the Riemann surfaces [Fig.~\ref{fig_topo}(b)]\footnote{For the $T=0.2 \ \mu$s loop, the control is limited by the 1~ns timing resolution of the experimental apparatus leading to a slight smearing of the response of $\xf$ at the EPs.}.   In Fig.~\ref{fig_topo}(c) we display the results for $T=1 \ \mu$s: $\xf$ exhibits similar features, yet with lower contrast  compared to $T=0.2 \ \mu$s. This is due to the emergence of decoherence effects as discussed previously.

\section{Conclusion and Outlook} \label{sec:conclusion}

We have demonstrated counterdiabatic driving to control a non-Hermitian dissipative quantum system. By implementing additional driving terms, we enabled the qubit to follow adiabatic paths in significantly shorter timescales while preserving the topological features of the complex energy landscape. This approach effectively mitigated the breakdown in adiabaticity that typically occurs when encircling EPs, allowing us to resolve the Riemann sheet topology with high fidelity. While our implementation has proven effective, several opportunities for future investigation are present. First, we identified specific parameter trajectories—Apollonius circles—that eliminate the anti-Hermitian components of $H_\textsc{cd}$. This enables us to approach the exceptional points more closely, where distinctive non-Hermitian features such as eigenstate non-orthogonality become pronounced, while still preserving Hermitian counterdiabatic control. 
This would allow us to explore the strong counterdiabatic limit, where interesting opportunities for novel control paradigms and quantum state manipulation arise. 
Second, we observed that for long evolution times, the efficacy of counterdiabatic driving diminishes due to decoherence effects. This represents a fundamental difference from classical non-Hermitian systems, where counterdiabatic driving remains effective even in the long-time limit. Future work could explore how these control techniques might be combined with decoherence mitigation strategies such as dynamical decoupling to extend the coherent control of non-Hermitian quantum systems over longer timescales. Finally, the methods introduced in this work were implemented for a system with EP2 degeneracies. However, extension of these methods to higher order degeneracies could offer insight into the non-Hermitian topology in larger systems~\cite{zhang2024,Wu2024,Chen2024}.
Overall, this work establishes counterdiabatic driving as a powerful tool for exploring and harnessing the unique topological features of non-Hermitian quantum systems, with potential applications in quantum sensing, state preparation, and geometric phase manipulation.

\section{Acknowledgments}
We are grateful to Jack Harris for discussions.
This research was supported by NSF Grant No.~PHY-2408932
and No.~2152221, 
the Air Force Office of Scientific Research (AFOSR) Multidisciplinary University Research Initiative (MURI) Award on Programmable systems with non-Hermitian quantum dynamics (Grant No. FA9550-21-1-0202), 
ONR Grant No.~N000142512160, 
the Institute of Materials Science and Engineering at Washington University, and the Luxembourg National Research Fund (FNR, Attract grant QOMPET 15382998). Devices were fabricated and provided by the Superconducting Qubits at Lincoln Laboratory (SQUILL) Foundry at MIT Lincoln Laboratory, with funding from the Laboratory for Physical Sciences (LPS) Qubit Collaboratory.

\begin{appendix}


\section{Mapping between Bloch-coordinates and parameter space\label{appA}}
We here relate the Bloch-coordinates of the eigenstates $|R_\pm\rangle$ to the parameter space coordinates and derive an expression of the detuning and coupling in terms of the rotational and hyperbolic angles, $\alpha_\textsc{r}$ and $\alpha_\textsc{i}$. We start with the case of real coupling before presenting the general case. 
\subsection{Real coupling ($\phi=0$)}
\begin{figure}
    \centering
    \includegraphics[width=.5  \textwidth]{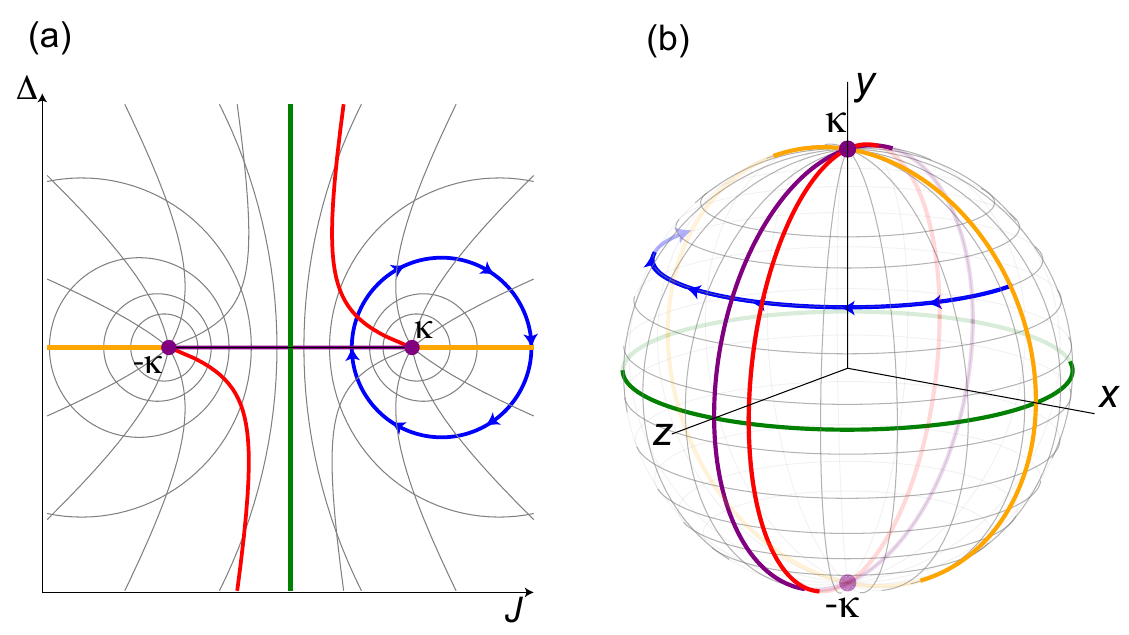}
    \caption{Illustration of the mapping between parameter coordinates (a) and Bloch coordinates (b). Each colored parameter trajectory covers half of the corresponding  Bloch space trajectory. For example, the encircling trajectory (blue), corresponds to tuning parameters around the EP at $\kappa$. In Bloch space, this trajectory wraps only half of the Bloch sphere.  }
    \label{figure: bloch map}
\end{figure}

The mapping between the eigenstates Bloch-coordinates $(z_\pm,x_\pm,y_\pm)$ and the complex angle $\alpha$ can be made explicit by separating $\alpha$ into a real angle $\alpha_\textsc{r}$ and hyperbolic angle $\alpha_\textsc{i}$. Explicitly evaluating the expectation values of the Pauli matrices yields,
\begin{equation}
\begin{split}
    \label{bloch-coordinates}
    z_\pm  &= \mp \cos\alpha_\textsc{r}\sqrt{1-\tanh^2\alpha_\textsc{i}},\\
    x_\pm &= \mp \sin\alpha_\textsc{r}\sqrt{1-\tanh^2\alpha_\textsc{i}},\\
    y_\pm &= \tanh\alpha_\textsc{i}.
\end{split}
\end{equation}
$(\alpha_\textsc{r},\alpha_\textsc{i})$ can be visualized as the coordinates specifying the surface of a sphere; the north and south poles (with respect to the $y$-axis) are mapped to the exceptional points $\pm \kappa$ in parameter space. The constant-$\alpha_\textsc{i}$ slices follow the Apollonius circles in parameter space. As shown in Fig.~\ref{figure: bloch map}, tracing one Apollonius circle corresponds to a half circling in the state space.  Constant-$\alpha_\textsc{r}$ lines (longitudes) are mapped to the equilateral hyperbolas that intersect the Apollonius circles.

In order to express the detuning and coupling in terms of the rotation angle and the hyperbolic angle, we begin with the expression for $\alpha$ in Eq.~\eqref{rotation angle} and its complex conjugate to solve for $\varepsilon$. This yields
\begin{equation}
    \varepsilon = i \kappa\frac{(e^{-2i\alpha}+e^{-2i\alpha^*}) - 2}{e^{-2i\alpha} - e^{-2i\alpha^*}} = i \kappa\frac{\cosh(2\alpha_\textsc{i}) - e^{2i\alpha_\textsc{r}}}{\sinh(2\alpha_\textsc{i})}.
\end{equation}
From this expression, we can extract the detuning and the coupling strength as
\begin{subequations}\label{J-Delta angle}
\begin{align}
    \Delta &= 2 \kappa \frac{\sin(2\alpha_\textsc{r})}{\sinh(2\alpha_\textsc{i})},\\
    |J| &= \kappa\frac{\cosh(2\alpha_\textsc{i}) - \cos(2\alpha_\textsc{r})}{\sinh(2\alpha_\textsc{i})}.
\end{align}
\end{subequations}
Note that the minimal and maximal value of $|J|$ on an Apollonius circle is given by $J_\textrm{min} = \kappa\tanh\alpha_\textsc{i}$ and $J_\textrm{max} = \frac{\kappa}{\tanh\alpha_\textsc{i}}$, respectively. Figure~\ref{figure: bloch map} displays the Apollonius circles in the $\Delta$ and $J$ parameter space as well as the mapping to Bloch coordinates.

\begin{figure}[b]
    \centering
    \includegraphics[width=0.65\linewidth]{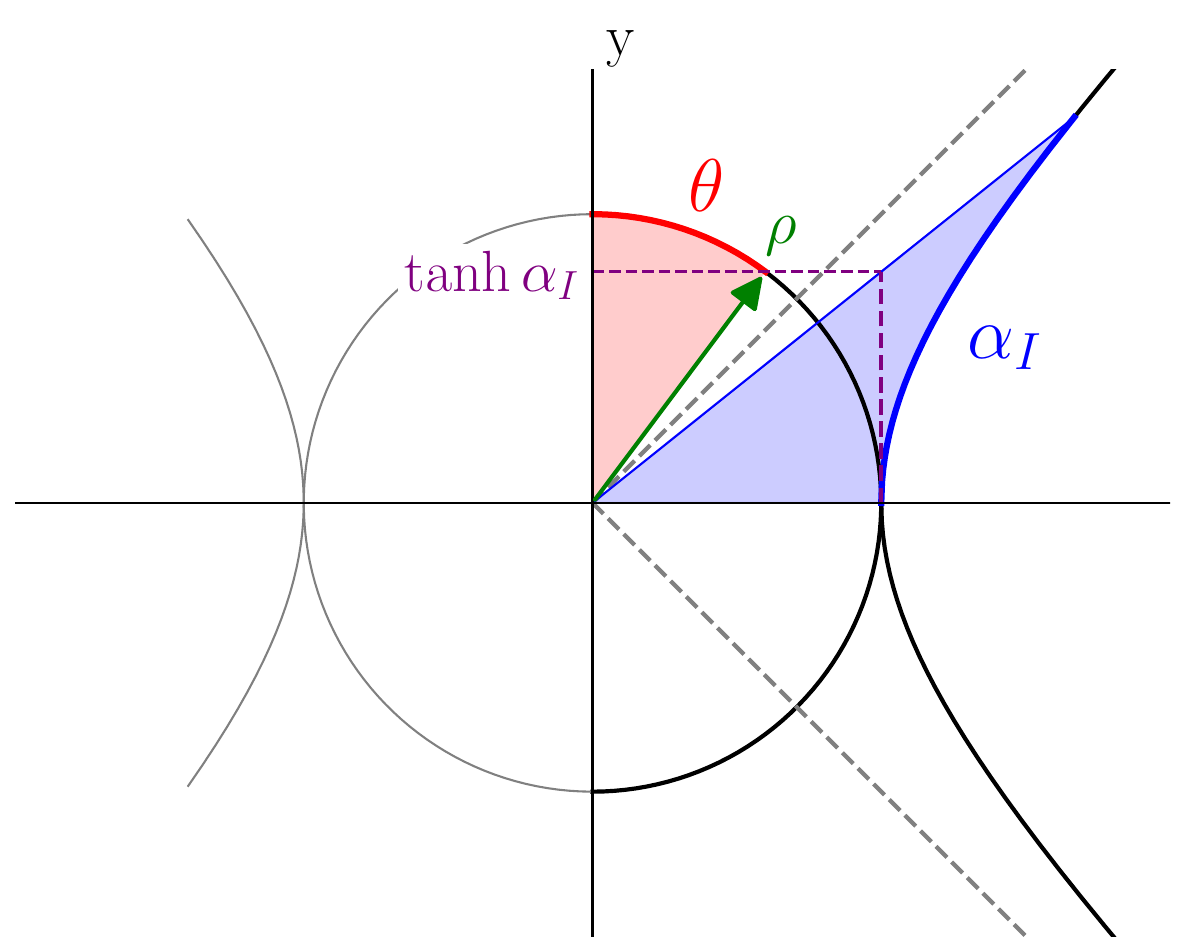}
    \caption{The figure shows the relationship between the angle $\theta$ between the right eigenvectors (red) and hyperbolic angle $\alpha_\textsc{i}$ with respect to the $y$-axis for a state $\rho$ on the Bloch sphere. The two angles are related via $\cos\theta = \tanh\alpha_\textsc{i}$ and the red and blue areas equals $\theta/2$ and $\alpha_\textsc{i}/2$ respectively.}
    \label{fig:hyperbolic}
\end{figure}

We can show that the angle $\theta$ between the right eigenvectors shares the same contour lines as the hyperbolic angle $\alpha_\textsc{i}$. Consequently, $\theta$ also remains constant along an Apollonius circle.
Indeed, from the overlap of the eigenvectors given in the main text, the transition probability between the normalized right eigenstates  
$|\Lambda_\pm\rangle$ reads
\begin{equation}
\label{overlap}
    |\langle \Lambda_- | \Lambda_+ \rangle|^2 = \frac{|\langle R_- | R_+ \rangle|^2}{\langle R_-| R_- \rangle\langle R_+| R_+ \rangle} = \tanh^2\alpha_\textsc{i}.
\end{equation}
Together with Eq.~\eqref{bloch-coordinates}, it follows that the angle between the two eigenstates is equal to the polar angle with respect to the $y$-axis, and given by $\theta\equiv \arccos|\tanh\alpha_\textsc{i}|$---see Fig.~\ref{fig:hyperbolic}. This overlap angle depends solely on \(\alpha_\textsc{i}\), so the contour lines of $\theta$ coincide with the Apollonius circles defined by the two exceptional points. This is intuitively expected: only the hyperbolic part of the rotation can alter the overlap between eigenstates, while the rotational part preserves it. This also explains why the counterdiabatic  Hamiltonian can remain Hermitian when $\alpha_\textsc{i}$ is constant, since this is precisely when the overlap angle is constant.

\subsection{Complex $J$ ($\phi \neq0$)}
We now extend the results from real to complex coupling $J = |J|e^{i\phi}$ by computing the mapping between the Bloch ball coordinates of the eigenstates $(z_\pm,x_\pm,y_\pm)$ and the three-dimensional parameter space $(\Delta, J_x, J_y)$. We will see that the Apollonius circles, corresponding to a fixed overlap between the eigenstates, generalize to toroidal surfaces, centered around the detuning axis.

As seen in Section \ref{sec:method} in the main text, the eigenvectors can then be expressed as a complex rotation of the computational basis followed by a rotation along the $z$-axis:
\begin{align}
\label{eigenstates general}
    |R_\pm\rangle &= \hat{R}_z(\phi)\hat{C}_y(\alpha)|z_\pm\rangle,
\end{align} 
This definition of $\alpha=\arctan\frac{|J|}{E}$ results in its imaginary part taking only positive values. It also ensures that $\alpha$ remains invariant  under the phase $\phi$ variation, corresponding to a rotation around the detuning-axis in parameter space. Under this variation, the Apollonius circles sweep out toroidal surfaces that enclose the exceptional ring. We refer to these as \textit{Apollonius tori}, each characterized by a constant hyperbolic angle $\alpha_\textsc{i}$. This parametrization is smooth in both $\alpha$ and $\phi$ for $|J|>0$ and remains single valued for $\,\alpha_\textsc{r} \in (-\pi/2, \pi/2]$. Due to Eq.~\eqref{overlap} and Eq.~\eqref{eigenstates general}, we also note that $\theta$ is independent of $\phi$ and therefore remains constant along an Apollonius torus.

The Bloch coordinates $(z_\pm,x_\pm,y_\pm)$ of the right eigenvectors take the form 
\begin{equation}
\begin{split}
    \label{bloch-coordinates general}
    z_\pm  &= \mp \cos\alpha_\textsc{r}\sqrt{1-\tanh^2\alpha_\textsc{i}},\\
    x_\pm &= \mp \cos\phi \sin \alpha_\textsc{r} \sqrt{1-\tanh^2\alpha_\textsc{i}}+ \sin\phi\tanh\alpha_\textsc{i},\\
    y_\pm &= \cos\phi \tanh\alpha_\textsc{i} \pm \sin\phi\sin\alpha_\textsc{r}\sqrt{1-\tanh^2\alpha_\textsc{i}},
\end{split}
\end{equation}
while the detuning and coupling modulus remain unchanged, as given in Eq.~\eqref{J-Delta angle}. 
 It is instructive to compute the ratio
\begin{equation}
\frac{|J|}{\Delta/2} =  \tan\alpha_\textsc{r} + 2\frac{\sinh^2\alpha_\textsc{i}}{\sin\alpha_\textsc{r}}.
\end{equation}
We observe that $\alpha_\textsc{r}$ reduces to the standard mixing angle when $\alpha_\textsc{i} = 0$. However, in the limit $\alpha_\textsc{i}\rightarrow 0$ with $\alpha_\textsc{r}\neq 0$, both $\Delta$ and $J$ diverge, indicating that an infinite amount of energy is required to rotate the eigenstates along the equator (with respect to the $y$-axis) when $\kappa >0$. Conversely, as we approach an exceptional point, the values of $\Delta$ and $J$ required to vary $\alpha_\textsc{r}$ decrease.

\section{Counterdiabatic Hamiltonian}
\label{Counterdiabatic_Hamiltonian_phi}
\subsection{General expression for $\phi\neq0$ }
We derive a counterdiabatic Hamiltonian for the general case of complex coupling ($\phi\neq0$). The expression we obtain ensures parallel transport when $\dot{\phi} = 0$.

The transport operator along the path $(\alpha_t,\phi_t)$,
\begin{equation}
    \hat{T}(t;t_0) \equiv \hat{R}_z(\phi_t)\hat{C}_y(\alpha_t)\hat{C}_y(-\alpha_{t_0})\hat{R}_z(-\phi_{t_0}),
\end{equation}
transports the eigenbasis Eqs.~\eqref{eigenstates general} from time $t_0$ to $t$. However, this no longer implements a parallel transport when $\phi$ varies with time, as the Berry connection becomes
\begin{equation}
\langle L_\pm|\partial_t R_\pm\rangle = \pm i\frac{\dot{\phi}}{2}\cos\alpha.
\end{equation}
Despite this, we can still define a counterdiabatic Hamiltonian that generates this transport: 
\begin{equation}
    H_\textsc{cd} = i\dot{\hat{T}}\hat{T}^{-1} = \frac{\dot{\phi}}{2}\hat{\sigma}_z + \frac{\dot{\alpha}}{2}\hat{R}_{z}(\phi)\hat{\sigma}_y \hat{R}^\dagger_{z}(\phi).
\end{equation}
As in the previous case, the anti-Hermitian component vanishes when $\alpha_{\textsc{i}}$ remains constant. This implies that the transport is Hermitian precisely when the trajectory in parameter space lies on an Apollonius torus defined by the exceptional ring.

Despite the failure of parallel transport, the total dynamical phase accumulated along certain special trajectories can vanish. For example, consider a trajectory on an Apollonius torus with constant angular velocities $\dot{\alpha}_{\textsc{r}} = \omega$ and $\dot{\phi} = \nu$. When the path winds twice around the minor radius of the torus, the dynamical phase accumulated over this cycle cancels out exactly. To see this, note that the parallel-transported eigenstate is given by
\begin{equation}
    |R^\parallel_\pm(t)\rangle = e^{-\int_{t_0}^t\langle L_\pm(s)|\partial_t R_\pm(s)\rangle \textrm{d}s}|R_\pm(t_0)\rangle.
\end{equation}
The accumulated dynamical phase vanishes as
\begin{align}
    &\int_{t_0}^t\langle L_\pm(s)|\partial_t R_\pm(s)\rangle\, \textrm{d}s=\pm \frac{i}{2}\int_{t_0}^t\nu\cos\alpha(s) \, \textrm{d}s\\
    &=\pm \frac{i\nu}{2\omega}\Bigg[\cosh(\alpha_\textsc{i})\int_{\alpha_\textsc{r}(0)}^{\alpha_\textsc{r}(0)+2\pi}\cos\alpha_\textsc{r} \textrm{d}\alpha_\textsc{r}\nonumber \\
    &\qquad \quad - i\sinh\alpha_\textsc{i}\int_{\alpha_\textsc{r}(0)}^{\alpha_\textsc{r}(0)+2\pi}\sin\alpha_\textsc{r} \, \textrm{d}\alpha_\textsc{r}\Bigg] \nonumber \\
    &= 0 \nonumber
\end{align}
Note that the parameter $\alpha_\textsc{r}$ is a multivalued angular coordinate, and its evolution along the closed path (e.g., from $\alpha_\textsc{r}(0)$ to $(\alpha_\textsc{r}(0) + 2\pi$)) will cross a branch cut. However, since the path is smooth, we can extend $\alpha_\textsc{r}(t)$ to a smooth, single-valued function along the trajectory. This corresponds to analytically continuing $\alpha_\textsc{r}$ along the path, ensuring that all derived quantities, such as $\alpha(t)$, remain smooth and well-defined. In this sense, the integral above remains well defined even when the trajectory loops around a branch point, as it correctly accounts for the Riemann sheet structure of the underlying functions.

\subsection{On the Limits of Hermitian Counterdiabatic Driving in Parallel Transport}
\label{Limits_Hermitian_CD}

Here we derive the parallel transporting counterdiabatic drive for the general case and its conditions for which it is Hermitian.

The transport operator that parallel transports the eigenbasis along the path $(\alpha_t,\phi_t)$ is given by $
    \hat{T}_\parallel(t;t_0) \equiv \hat{R}_z(\phi_t)\hat{C}_y(\alpha_t)\hat{R}_z(\beta_t)\hat{C}_y(-\alpha_{t_0})\hat{R}_z(-\phi_{t_0})\hat{R}_z(-\beta_0)$,
where
\begin{equation}
    \beta(t) = -\int_{t_0}^t \dot{\phi}(s)\cos{\alpha(s)\,\textrm{d}s}.
\end{equation}
The corresponding counterdiabatic Hamiltonian takes the form
    $H_\textsc{cd}^\parallel = i\dot{T}_\parallel T^{-1}_\parallel
    =\frac{\dot{\phi}}{2}\hat{\sigma}_z + \frac{\dot{\alpha}}{2}\hat{R}_z(\phi)\hat{\sigma}_y \hat{R}^\dagger_z(\phi) - \frac{\dot{\phi}}{2}\cos\alpha\,\hat{R}_{z}(\phi)\hat{C}_{y}(\alpha)\hat{\sigma}_z \hat{C}_{y}^{-1}(\alpha)\hat{R}_{z}^\dagger(\phi).$
The anti-Hermitian part of this expression vanishes if and only if
\begin{equation}
    \dot{\alpha}\hat{\sigma}_y - \dot{\phi}\cos\alpha\,\hat{\sigma}_z = 0,
\end{equation}
which occurs precisely when $\mathrm{Im}( \dot{\alpha}) = 0$ (i.e., constant $\alpha_\textsc{i}$), and either $\dot{\phi} = 0$ or $\cos{\alpha} = 0$. The second condition corresponds to $\alpha_\textsc{i} = 0$, or equivalently $|J| = 0$.

\section{Chiral symmetry} \label{appC}

The existence of Hermitian-only counterdiabatic controls, as exemplified by Apollonius circles, relies on the ability to parameterize the state in terms of a rotation angle and a hyperbolic angle relative to a fixed axis on the Bloch sphere. In this Appendix we show that this stems from an underlying chiral symmetry of the Hamiltonian.

To make this symmetry manifest, we shift the Hamiltonian spectrum such that it becomes traceless, defining
\begin{equation}
    H' \equiv H - \frac{1}{2}\textrm{Tr}(H)\, \hat{I}.
\end{equation}
The shifted Hamiltonian $H'$, which shares the same eigenvectors $|R_\pm\rangle$ as $H$, then satisfies a chiral symmetry relation of the form
\begin{equation}
    \hat{\Gamma} H' \hat{\Gamma} = -H', \quad \hat{\Gamma}^2 = \hat{I},
\end{equation}
where the chiral symmetry operator is given by
\begin{equation}
\label{chiral operator}
    \hat{\Gamma} = \hat{R}_z(\phi)\, \hat{\sigma}_y\, \hat{R}_z(\phi)^\dagger.
\end{equation}
This symmetry exchanges the eigenstates: $\hat{\Gamma}|R_\pm\rangle = \pm i|R_\mp\rangle$.

Away from exceptional points, the eigenvectors $\{|R_+\rangle, |R_-\rangle\}$ form a basis and can be related to the basis $\{|z_+\rangle, |z_-\rangle\}$ via an invertible transformation:
\begin{equation}
\label{special linear}
    |R_\pm(\Delta, J)\rangle = \hat{G}(\Delta, J) \hat{R}_z(\phi)\, |z_\pm\rangle, \quad \hat{G}(\Delta, J) \in \mathrm{SL}(2, \mathbb{C}).
\end{equation}
Combining the relations $\hat{\Gamma}|R_\pm(\Delta, J)\rangle = \pm i|R_\mp(\Delta, J)\rangle$ and $\hat{\Gamma} \hat{R}_z(\phi)\,|z_\pm\rangle = \pm i\hat{R}_z(\phi)\,|z_\mp\rangle$ from \eqref{chiral operator} implies that the transformation $\hat{G}(\Delta, J)$ commutes with $\hat{\Gamma}$, i.e., $[\hat{G}, \hat{\Gamma}] = 0$.

Since any element of the special linear group $\mathrm{SL}(2, \mathbb{C})$ can be written as an exponential $\hat{G} = e^{\hat{A}}$ with $\mathrm{Tr}(\hat{A}) = 0$, the commutation relation $[\hat{G}, \hat{\Gamma}] = 0$ implies that $\hat{A}$ must also commute with $\hat{\Gamma}$. This constraint forces $\hat{A}$ to be proportional to $\hat{\Gamma}$, and we can write
\begin{equation}
\label{G generator}
    \hat{A} = -i \alpha(\Delta, J)\, \hat{\Gamma}
\end{equation}
for some complex function $\alpha$ over parameter space. It follows that
\begin{equation}
\label{G as complex rotation}
    \hat{G}(\Delta, J) = e^{-i\frac{\alpha(\Delta, J)}{2}\hat{\Gamma}} = \hat{R}_z(\phi)\, e^{-i\frac{\alpha(\Delta, J)}{2}\hat{\sigma}_y}\, \hat{R}_z(\phi)^\dagger,
\end{equation}
and together with Eq.~\eqref{special linear} we obtain
\begin{equation}
    |R_\pm\rangle = \hat{R}_z(\phi)\, \hat{C}_y(\alpha)\, |z_\pm\rangle,
\end{equation}
where $\hat{C}_y(\alpha) = e^{-i\frac{\alpha}{2}\hat{\sigma}_y}$ describes the complex rotation about the $y$-axis.

We may regard Eq.~\eqref{G generator} as the defining property of $\alpha$. Letting $\pm\xi$ denote the eigenvalues of $H'$, the spectral theorem, together with Eq.~\eqref{special linear} and Eq.~\eqref{G as complex rotation} gives 
\begin{align}
    E &= \langle z_+|\hat{R}_z(\phi)^\dagger H'\hat{R}_z(\phi)|z_+\rangle = \xi\cos\alpha,\\
    |J| &= \langle z_+|\hat{R}_z(\phi)^\dagger H'\hat{R}_z(\phi)|z_-\rangle = \xi\sin\alpha,
\end{align} 
from which the identity $\tan \alpha = \frac{|J|}{E}$ immediately follows.

\end{appendix}



%

\end{document}